\documentclass[twocolumn,showpacs,amssymb,aps,prc,nofootinbib,floatfix,superscriptaddress]{revtex4-1}
\usepackage{amsmath,graphicx,color,ulem}
\usepackage{hyperref}

\newcommand{\nn}{{\nonumber}}
\def \beq{\begin{equation}}
\def \eeq{\end{equation}}
\def \beqa{\begin{eqnarray}}
\def \eeqa{\end{eqnarray}}

\def \sNN{$\sqrt{s_{NN}}$ }
\def \spt{[p_T] }
\def \mpt{\langle p_T \rangle }
\def \dpt{\delta p_T }

\begin{document}
\title{Rapidity dependence of mean transverse momentum fluctuation and decorrelation in a baryon-dense medium}

\author{Tribhuban Parida}
\affiliation{AGH University of Krakow, Faculty of Physics and Applied Computer Science, aleja Mickiewicza 30, 30-059 Cracow, Poland}

\begin{abstract} 
 I study the event-by-event fluctuation and rapidity decorrelation of the mean transverse momentum $\spt$, which has recently been proposed as a sensitive probe of the equation of state at finite baryon density. The investigation reveals that, in a baryon-rich medium, the event-by-event fluctuation of the mean transverse momentum is driven by the combined effects of energy-density and net-baryon-density fluctuations. Consequently, the rapidity dependence of this observable provides a promising handle to probe the three-dimensional structure of both energy and baryon density profiles. Previous studies have shown that $\spt$ decorrelation along rapidity is largely insensitive to shear and bulk viscosity; however, its dependence on baryon diffusion, another key transport coefficient in baryonic matter, has not been explored. I find that baryon diffusion has a negligible impact, establishing this observable as a robust probe of the equation of state. Furthermore, I present predictions for identified hadrons and observe a pronounced splitting in the rapidity decorrelation of mean transverse momentum between protons and antiprotons, indicating different transverse flow dynamics for baryons and antibaryons.
\end{abstract}

\maketitle

\section{Introduction}
In non-central relativistic heavy-ion collisions, the initially deposited matter occupies an almond-shaped region in the transverse plane. The subsequent hydrodynamic expansion is driven by anisotropic pressure gradients, which convert the initial spatial anisotropy into a momentum-space anisotropy of the produced particles~\cite{Ollitrault:1992bk,Ollitrault:2007du}. This process, known as shape to flow transmutation~\cite{Bozek:2014cva}, is controlled by the properties of the medium, including its transport coefficients and equation of state~\cite{Song:2010mg,Bozek:2009dw,Bozek:2011ua,Karpenko:2015xea,Shen:2015msa,Ryu:2015vwa,Shen:2020jwv,OmanaKuttan:2022aml,Gong:2024lhq}. Over the past years, systematic measurements of anisotropic flow coefficients across a wide range of collision energies~\cite{STAR:2000ekf,ALICE:2011ab,PHENIX:2003qra,STAR:2015rxv}, together with detailed comparisons between theoretical models and experimental data, have led to substantial progress in constraining the properties of the strongly interacting QCD medium \cite{Bernhard:2016tnd,Bernhard:2019bmu,Nijs:2020ors,JETSCAPE:2020mzn,Jahan:2024wpj}.

While anisotropic flow has been studied extensively, comparatively less studies has been devoted to radial flow and its phenomenological consequences \cite{Schenke:2020uqq,Samanta:2025yrj,Parida:2024ckk,Samanta:2023amp,Jia:2025rab,Bhatta:2025oyp,Zhou:2025bwu,Giacalone:2020lbm,Gardim:2019brr,Saha:2025nyu}. In a baryon-free medium, which is approximately realized near mid-rapidity at LHC energies, event-by-event fluctuations in the initial transverse size give rise to variations in the initial energy density \cite{Samanta:2023amp,Bozek:2017elk,Gardim:2019brr,Parida:2024ckk}. For events with similar multiplicity, a smaller initial size corresponds to a compact energy deposition, leading to stronger radial pressure gradients and consequently a more rapid transverse expansion \cite{Broniowski:2009fm}. This enhanced expansion results in the final state as an increase in the transverse momentum per produced particle, or equivalently, an increase in the mean transverse momentum $\spt$. This mechanism is commonly referred to as size to flow transmutation \cite{Bozek:2017elk}. Since the pressure response to a given change in energy density is governed by the equation of state, size to flow transmutation provides direct sensitivity to the EoS of the medium \cite{Gong:2024lhq,Mu:2025gtr,Gardim:2019xjs}. Motivated by this connection, studies at top RHIC and LHC energies have explored event-by-event fluctuations of $\spt$ to extract information on the speed of sound of the QCD medium \cite{Gardim:2019xjs,Gardim:2019brr,Gardim:2019brr,CMS:2023byu}. Moreover, in the context of radial flow, a recently proposed observable, denoted as $v_0$ and its $p_T$-differential counterpart $v_0(p_T)$, quantifies radial flow through correlations between the mean transverse momentum and the spectra \cite{Schenke:2020uqq,Parida:2024ckk, Du:2025hrz, ATLAS:2025ztg, ALICE:2025iud}, and is expected to be sensitive to the bulk viscosity of the medium \cite{Parida:2024ckk,Du:2025dpu}.

The situation becomes qualitatively different in the baryon-rich matter created in the RHIC Beam Energy Scan program. In the presence of a finite net-baryon density, the pressure ($\mathcal{P}$) depends not only on the energy density ($\epsilon$) but also on the net-baryon density ($\rho_B$) \cite{Ratti:2018ksb,Monnai:2019hkn,Monnai:2024pvy,Noronha-Hostler:2019ayj,Mondal:2021jxk}. As a result, event-by-event $\mathcal{P}$ fluctuations receive contributions from fluctuations in both quantities. The additional baryonic degree of freedom therefore enriches the dynamics of transverse expansion and makes the study of transverse flow fluctuations in a baryon-dense medium particularly more informative.

At top RHIC and LHC energies, heavy-ion collisions approximately exhibit longitudinal boost invariance \cite{Bjorken:1982qr,Dumitru:2008wn}, and as a consequence, studies of transverse flow are often restricted to the mid-rapidity region. Most hydrodynamic calculations at these energies therefore assume boost invariance and employ a (2+1)D hydrodynamic evolution when investigating transverse flow and it's fluctuations \cite{Song:2009gc,Noronha-Hostler:2013gga,Luzum:2009ue,Samanta:2025fuj,Gelis:2019vzt,Bhalerao:2019fzp, Giacalone:2020ymy,Bernhard:2018hnz,Roy:2012jb,Song:2011hk}. Nevertheless, there exist studies that relax this assumption and explore flow fluctuations and correlations along the longitudinal direction even at LHC energies \cite{Pang:2015zrq,Pang:2014pxa, Bozek:2010vz, Bzdak:2012tp, Xiao:2012uw, Huo:2013qma, Jia:2014ysa, CMS:2015xmx, ATLAS:2017rij, ATLAS:2020sgl, Bozek:2015bna, Bozek:2015bha, Chatterjee:2017mhc}. In contrast, at lower collision energies, boost invariance is strongly broken \cite{STAR:2014clz, STAR:2019vcp, Nie:2020trj, Cimerman:2021gwf}. Experimental measurements of charged particle multiplicity and net-proton rapidity distributions indicate pronounced rapidity dependence in both the $\epsilon$ and $\rho_B$ profiles \cite{Back:2002wb,NA49:1998gaz,BRAHMS:2003wwg}. This makes it essential to investigate transverse flow fluctuations and their correlations across rapidity. Such measurements can provide valuable insight into the initial three-dimensional structure of the fireball and the resulting longitudinal dynamics of the system.

As discussed above, radial flow fluctuations are driven by pressure fluctuations in the medium. For given fluctuations in $\epsilon$ and $\rho_B$, the equation of state determines how these variations are converted into pressure fluctuations, which in turn govern the transverse expansion. Since both the $\epsilon$ and $\rho_B$ vary significantly along rapidity, the pressure and consequently the $\spt$ also fluctuate in a rapidity-dependent manner. The study of $\spt$ fluctuations as a function of rapidity, together with their correlations across rapidity, therefore provides a sensitive probe of the equation of state of the QCD medium at finite baryon density \cite{Liu:2025fbu}. In addition to the equation of state, transport coefficients play an important role in shaping the transverse expansion dynamics. In particular, bulk viscosity modifies the effective pressure and influences the rate of transverse expansion \cite{Bozek:2009dw, Ryu:2015vwa, Noronha-Hostler:2013gga}. Measurements of rapidity-dependent $\spt$ fluctuations and their longitudinal correlations thus may offer an opportunity to constrain the temperature and baryon chemical potential dependence of the transport properties of the QCD medium \cite{Shen:2020jwv,Jahan:2024wpj}.

For these reasons, the rapidity decorrelation of $\spt$ emerges as a particularly promising observable. Recent studies have demonstrated that $\spt$ fluctuations along rapidity are sensitive to the equation of state and have the potential to constrain the EoS of baryon-rich matter \cite{Liu:2025fbu}. This motivates the present work, in which I investigate the origin of $\spt$ fluctuation and it's rapidity decorrelation in a baryonic medium.

The primary objective of this paper is to identify the mechanisms that drive transverse momentum fluctuations in a baryon-dense fireball. Unlike the baryon-free case, where initial size fluctuations dominate \cite{Samanta:2023amp,ATLAS:2022dov}, $\spt$ fluctuations in a baryonic medium arise from the combined effects of energy density and net-baryon density fluctuations. I demonstrate the role of these two sources. Furthermore, I study the impact of baryon diffusion on $\spt$ rapidity decorrelation. I extend the analysis to identified hadrons and show that, in a baryon-dense medium, $\spt$ fluctuation and it's rapidity decorrelation exhibits a clear splitting between protons and antiprotons.

The remainder of this paper is organized as follows. In the next section, I introduce the observables studied in this work and present their definitions. Section~\ref{sec:model} describes the theoretical framework and the model employed in this study. The results are presented and discussed in Section~\ref{sec:results}. Finally, Section~\ref{sec:summary} summarizes the main observations and conclusions.

\section{Definitions}
\label{sec:definitions}
For a given event, the mean transverse momentum per particle within a pseudorapidity window $\eta$ is denoted by $\spt_\eta$ and is defined as
\beq
\spt_\eta = \frac{1}{N_\eta} \sum_{i=1}^{N_\eta} (p_T)_i ,
\eeq
where $(p_T)_i$ is the transverse momentum of the $i$th particle and the sum runs over all particles within the chosen pseudorapidity interval. Here, $N_\eta$ denotes the total number of particles in that window for a given event.

The event-averaged mean transverse momentum at pseudorapidity $\eta$ is denoted by $\mpt_\eta$ and is obtained by averaging $\spt_\eta$ over all events,
\beq
\mpt_\eta = \frac{1}{N_{\rm ev}} \sum_{j=1}^{N_{\rm ev}} \spt_{\eta,j},
\eeq
where $N_{\rm ev}$ is the total number of events and $\spt_{\eta,j}$ is the mean transverse momentum in the $j$th event.

Event-by-event fluctuations of the mean transverse momentum are quantified by the deviation of $\spt_\eta$ from its event-averaged value,
\beq
\spt_\eta = \mpt_\eta + (\delta p_T)_\eta .
\eeq
The magnitude of these fluctuations is characterized by the root-mean-square (RMS) of $(\delta p_T)_\eta$,
\beq
(\sigma_{p_T})_\eta = \sqrt{ \left\langle (\delta p_T)_\eta^2 \right\rangle },
\eeq
where $\langle \cdots \rangle$ denotes averaging over all events.
Following Ref.~\cite{Schenke:2020uqq}, I define
\beq
v_0(\eta) = \frac{(\sigma_{p_T})_\eta}{\mpt_\eta} ,
\eeq

The event-by-event correlation of $\spt_\eta$ between two pseudorapidity or rapidity windows can be quantified using the Pearson correlation coefficient. Taking $\eta_{\rm ref}$ as a reference window, the correlation observable is defined as
\beq
R_{p_T}(\eta;\eta_{\rm ref}) =
\frac{\mathcal{C}(\spt_{\eta_{\rm ref}}, \spt_\eta)}
     {(\sigma_{p_T})_{\eta_{\rm ref}}\,(\sigma_{p_T})_\eta},
\label{def:Rpt}
\eeq
where $\mathcal{C}(\spt_{\eta_{\rm ref}}, \spt_\eta)$ denotes the covariance between the event-wise mean transverse momenta in the two pseudorapidity windows.

Correlations of $\spt_\eta$ across pseudorapidity reflect correlations of transverse flow at different longitudinal positions. To suppress non-flow contributions, one can exploit large pseudorapidity separations. In this context, a three-bin decorrelation observable is constructed as \cite{Chatterjee:2017mhc}
\beq
r_{p_T}(\eta;\eta_{\rm ref}) = \frac{\mathcal{C}(\spt_{\eta_{\rm ref}}, \spt_{-\eta})}
     {\mathcal{C}(\spt_{\eta_{\rm ref}}, \spt_\eta)} = 
\frac{R_{p_T}(-\eta;\eta_{\rm ref})}
     {R_{p_T}(\eta;\eta_{\rm ref})},
     \label{def:rpt}
\eeq
which provides a measure of the longitudinal decorrelation of transverse momentum fluctuations while further reducing the sensitivity to non-flow effects. In the above expression, the second equality is only true for symmetric collisions.

When studying the rapidity dependence of $\spt$ fluctuations and correlations for identified hadrons, I use the rapidity $y$ instead of the pseudorapidity $\eta$. Accordingly, in those cases, $\eta$ is replaced by $y$ in all the above definitions and expressions.

\section{Model description}
\label{sec:model}
In this work, I employ a hybrid framework to simulate relativistic heavy-ion collisions. The evolution of the system in the dense and high-temperature phase is described using relativistic viscous hydrodynamics, while the late-stage dilute hadronic phase is modeled using a microscopic hadronic transport approach. Specifically, the hydrodynamic evolution is performed using the publicly available code MUSIC \cite{Schenke:2010nt,Schenke:2011bn,Denicol:2018wdp}, and the hadronic afterburner stage is simulated with UrQMD \cite{Bass:1998ca,Bleicher:1999xi}. The transition from the fluid description to particle degrees of freedom is implemented via the Cooper--Frye prescription at a constant switching energy density $\epsilon_{\text{SW}}$, using the iSS sampler \cite{Shen:2014vra}. In the present study, this conversion is carried out at $\epsilon_{\text{SW}} = 0.26$~GeV/fm$^{3}$.

The MUSIC framework solves the conservation equations for both the energy--momentum tensor and the net-baryon current. The complete set of evolution equations for the conserved quantities as well as the dissipative currents at finite baryon density is described in detail in Ref.~\cite{Denicol:2018wdp}. In the simulations, I take a constant specific shear viscosity, $\eta T / (\epsilon + \mathcal{P}) = 0.08$. The bulk viscosity $\zeta$ is taken to be temperature dependent, following the same functional form as in Ref.~\cite{Schenke:2020mbo},
\beq
\frac{\zeta T}{\epsilon + \mathcal{P}} = B_\zeta \exp \left[ -\frac{(T-T_0)^2}{\sigma_\zeta^2} \right],
\eeq
The parameters are chosen as $B_\zeta = 0.13$, $T_0 = 0.16$~GeV, and $\sigma_\zeta = 0.01$~GeV for $T < T_0$, while $\sigma_\zeta = 0.12$~GeV is used for $T > T_0$. The baryon diffusion coefficient $\kappa_B$ is taken to be dependent on both the temperature $T$ and the baryon chemical potential $\mu_B$, following the parametrization given in Ref.~\cite{Denicol:2018wdp},
\beq
\kappa_{B} = \frac{C_B}{T} \rho_{B} \left[ \frac{1}{3} \coth{\left(\frac{\mu_B}{T}\right)} - \frac{\rho_B T}{\epsilon + \mathcal{P}} \right],
\eeq
where $\rho_B$ is the net-baryon density. The dimensionless parameter $C_B$ controls the overall strength of baryon diffusion and is treated as a tuning parameter in this study. For the equation of state, I employ the NEoS-BQS equation of state \cite{Monnai:2019hkn}, which enforces the local constraints $\rho_S = 0$ and $\rho_Q = 0.4\rho_B$. As a consequence, the system develops nonzero strangeness and electric charge chemical potentials during the hydrodynamic evolution.

A crucial input to the hydrodynamic simulation is the three-dimensional initial condition (IC) for the energy density and net-baryon density. For the initial energy density profile, I adopt the tilted fireball model \cite{Bozek:2010bi}. The functional form of the energy density distribution at a constant initial proper time $\tau_0$ is taken to be the same as in Ref.~\cite{Parida:2022lmt}. This parametrization generates a tilted profile in the reaction plane, where the magnitude of the tilt is controlled by a free parameter. For suitable choices of this parameter, the model successfully reproduces the negative mid-rapidity slope of the charged-hadron directed flow observed at top RHIC and LHC energies \cite{Bozek:2010bi,STAR:2008jgm,ALICE:2013xri}. The hydrodynamic evolution is initiated at a constant proper time $\tau_0$, and the initial flow velocity is assumed to follow Bjorken scaling. Accordingly, the initial four-velocity at each spacetime point $(x, y, \eta_s)$ is given by
\begin{equation}
u^\mu = (\cosh \eta_s, 0, 0, \sinh \eta_s).
\end{equation}

The initial net-baryon density distribution is taken from the parametrization proposed in Ref.~\cite{Parida:2022ppj,Parida:2025lhn},
\beqa 
\rho_{B} \left( x, y, \eta_s \right) &=& N_{B} \left[ \left( N_{+}(x,y) f_{+}^{n_B}(\eta_{s}) + N_{-}(x,y)f_{-}^{n_B}(\eta_{s}) \right)\right.\nn\\ &&\left.\times \left( 1- \omega \right) + N_\mathrm{coll} (x,y) f^{n_B}_{\mathrm{coll}}\left(\eta_{s}\right) \omega \right] \label{eq:nb3d} 
\eeqa
where $N_{+}(x,y)$ and $N_{-}(x,y)$ denote the transverse densities of participant nucleons from the forward- and backward-moving nuclei, respectively, and $N_\mathrm{coll}(x,y)$ represents the contribution from binary collisions. The rapidity-odd envelope functions $f_{+}^{n_B}(\eta_s)$ and $f_{-}^{n_B}(\eta_s)$ are taken to be same as used in Refs.~\cite{Denicol:2018wdp,Shen:2020jwv}. In addition, a symmetric rapidity component associated with binary collisions is included, with the form
\beq
f_{\mathrm{coll}}^{n_B}(\eta_s) = f_{+}^{n_B}(\eta_s) + f_{-}^{n_B}(\eta_s).
\label{eq:fcoll_etas}
\eeq

The normalization factor $N_B$ in Eq.~\eqref{eq:nb3d} is not treated as a free parameter. Instead, it is fixed by requiring that the total initially deposited net baryon number equals the number of participant nucleons,
\begin{equation}
\int \tau_{0}  d\eta_s  d^2r  \rho_{B}(x, y, \eta_s; \tau_{0}) = N_{\text{part}},
\label{net_baryon_is_npart_2}
\end{equation}
where $d^2r = dxdy$. Unlike more commonly used IC, where the transverse baryon deposition is determined solely by participant nucleons, the present parametrization incorporates a mixture of participant and binary-collision contributions. Introducing a nonzero value of $\omega$ modifies the transverse baryon distribution and has been shown to improve the description of the directed flow of identified hadrons, as well as the splitting between proton and antiproton $v_1$ \cite{Parida:2022zse,Parida:2022ppj,Parida:2025lhn}. While this particular choice of baryon IC is not essential for the qualitative conclusions of this work, it provides a realistic framework to study the influence of baryon density on the observables of interest. Alternative choices, such as purely participant-driven baryon deposition ($\omega = 0$), are not expected to alter the qualitative physics discussed in this paper. This expectation is explicitly verified in Appendix~\ref{app:omega_zero}, where I study the effect of $\omega$ on the rapidity dependence of the $\spt$ fluctuation.

\section{Results}
\label{sec:results}
I performed simulations for 0–10\% centrality Au+Au collisions at $\sqrt{s_{NN}} = 19.6$ GeV. The IC parameters were chosen to reproduce the pseudorapidity($\eta$) distributions of charged-particle multiplicity and rapidity dependence of net-proton yield. The experimental measurements of the net-proton rapidity distribution are not available at $\sqrt{s_{NN}} = 19.6$ GeV. Therefore, the model parameters were tuned to reproduce the rapidity differential net-proton yield measured in Pb+Pb collisions at $\sqrt{s_{NN}} = 17.3$ GeV by the NA49 Collaboration \cite{NA49:2010lhg}. But the availability of midrapidity net-proton measurement from STAR collaboration \cite{STAR:2017sal} provides further constraint on the rapidity profile of the baryon density.

Following the strategy adopted in Ref. \cite{Parida:2024ckk}, I investigate the mechanisms driving $\spt$ fluctuations, extending the analysis as a function of rapidity in a baryon-rich medium. Within this framework, I perform two types of simulations: one employing event-by-event fluctuating IC and the other using smooth (event-averaged) IC.

\begin{figure}[th!]
    \includegraphics[width=\linewidth]{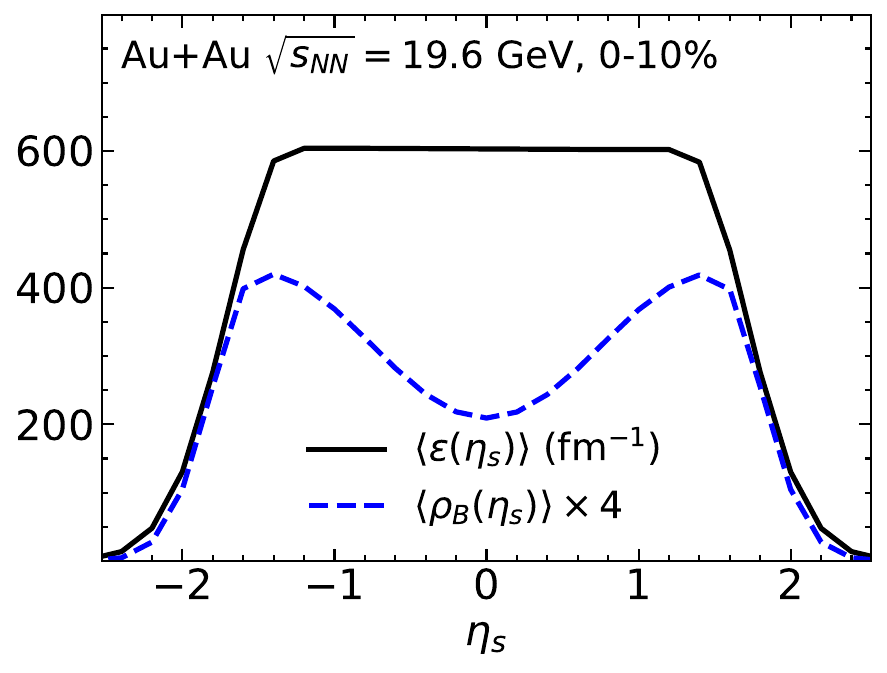}
    \caption{Space–time rapidity ($\eta_s$) dependence of the transverse-plane–integrated energy density and net-baryon density for 0–10\% central Au+Au collisions at $\sqrt{s_{NN}} = 19.6$ GeV.} 
    \label{fig:1}
\end{figure}

For the fluctuating IC scenario, I determine the centrality using the initial total transverse energy at mid-rapidity as an estimator, and select 200 independent initial configurations within the 0--10\% centrality class for hydrodynamic evolution. The space–time rapidity ($\eta_s$) dependence of the transverse-plane–integrated, event-averaged energy density and net-baryon density is shown in Fig. \ref{fig:1}. The energy density profile exhibits a broad plateau around midrapidity followed by a Gaussian fall-off at forward and backward rapidities, which allows to capture the experimental data of the charged-particle pseudorapidity distribution. In contrast, the net-baryon density displays a dip at midrapidity with two symmetric peaks located at finite rapidities. Consequently, the baryon density increases with increasing $|\eta_s|$ away from midrapidity. This parametrization enables the model to describe the experimental data of rapidity distribution of the net-proton yield.

Using these fluctuating IC, I perform event-by-event hydrodynamic evolution followed by hadronic transport. From the produced hadrons, I calculate the event-by-event mean transverse momentum $\spt$ and subsequently determine its fluctuation at each rapidity. To suppress statistical fluctuation, I oversample 1600 hadronic events from the particlization hypersurface corresponding to a given hydrodynamic evolution.

In the second scenario, I construct smooth initial profiles of the energy density and net-baryon density by averaging over $10^4$ initial-state configurations. These smooth profiles are then evolved using single-shot hydrodynamics. In this case, I consider four different variations:
\begin{enumerate}
\item a baseline case with the same normalization of energy density and net-baryon density as in the fluctuating IC scenario;
\item an increased energy density ($\epsilon$) normalization by a factor of 1.1 while keeping the net-baryon density unchanged;
\item an increased net-baryon density ($\rho_B$) normalization by a factor of 1.1 while keeping the energy density unchanged;
\item a simultaneous increase of both energy density ($\epsilon$) and net-baryon density ($\rho_B$) normalizations by a factor of 1.1.
\end{enumerate}

\begin{figure}[th!]
    \includegraphics[width=\linewidth]{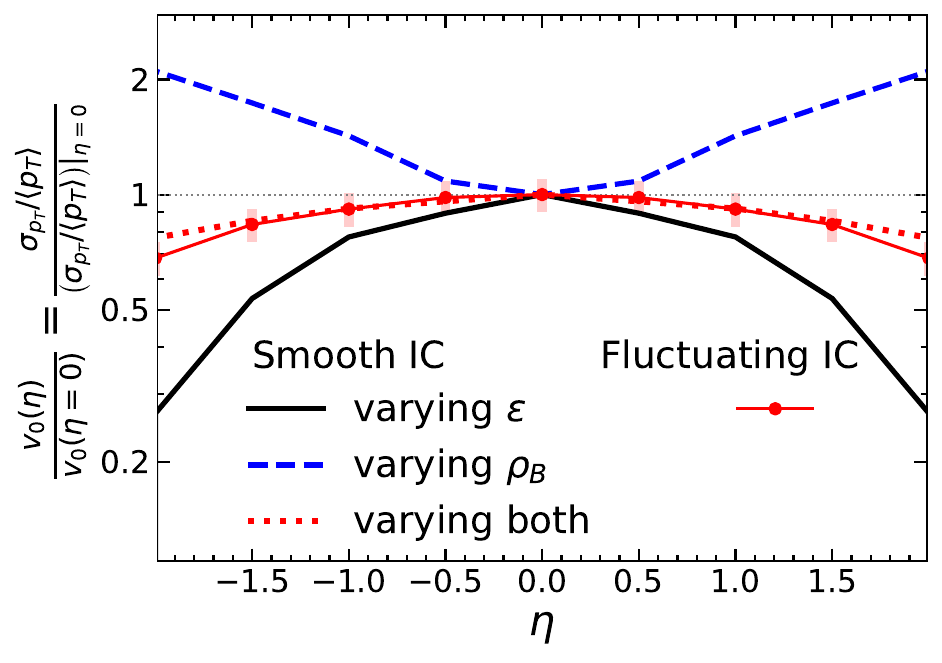}
    \caption{ Pseudo-rapidity dependence of the mean-$p_T$ fluctuation $v_0(\eta)/v_0(\eta=0)$, scaled by its respective midrapidity value, for smooth and fluctuating IC. The comparison illustrates the competing roles of energy-density and net-baryon-density fluctuations in driving rapidity dependence.} 
    \label{fig:2}
\end{figure}

Single-shot hydrodynamic evolution is performed for all four cases. In the last three scenarios, variations in $\spt$ are observed at each rapidity relative to the baseline case. From these deviations, I extract the corresponding fluctuation, $\sigma_{p_T} = \vert \dpt \vert$ and evaluate the relative change as
\begin{equation}
\frac{\delta p_T}{\mpt} = \delta \ln(\spt) = \ln\left( \frac{\spt_v}{\spt_b} \right),
\end{equation}
where $\spt_v$ and $\spt_b$ denote the mean transverse momentum obtained in the varied and baseline cases, respectively.

It should be emphasized that the normalization of the energy and net-baryon densities is varied arbitrarily by 10\% in the smooth IC study. Consequently, a direct quantitative comparison of $\delta p_T / \mpt$ between the fluctuating and smooth IC scenarios does not yield a meaningful conclusion. Instead, since my primary interest lies in the rapidity dependence of mean transverse momentum fluctuations, I analyze the rapidity dependence of $\delta p_T / \mpt$ after scaling it by its corresponding midrapidity value.

I observe that, when only the energy density normalization is varied, the resulting mean transverse momentum fluctuation decreases with increasing rapidity. This behavior can be understood from the $\eta_s$ dependence of the energy density profile shown in Fig. \ref{fig:2}. Around midrapidity, where the energy density is highest, a fixed 10\% variation in normalization leads to a relatively larger change in the local temperature, which in turn produces a stronger variation in $\spt$. At larger non-zero rapidities, the energy density is comparatively lower; consequently, the same fractional change in energy density induces a smaller temperature variation. As a result, the corresponding change in $\spt$ is reduced away from midrapidity, leading to the observed monotonic decrease of $\spt$ fluctuations with rapidity.

In contrast, varying the net-baryon density normalization leads to an opposite rapidity dependence: the $\spt$ fluctuation is minimal at midrapidity and increases toward larger rapidities. This trend directly reflects the longitudinal structure of the baryon density distribution, which exhibits a dip at midrapidity and increases toward finite rapidities, as shown in Fig. \ref{fig:1}. In a baryon-rich medium, temperature depend on both local energy density and the net-baryon density. Therefore, when the energy density is held fixed, a variation in baryon density produces a relatively small temperature change near midrapidity but a significantly larger change at forward and backward rapidities, where the baryon density is higher. This enhanced temperature sensitivity away from midrapidity leads to a stronger variation of $\spt$ in those regions, resulting in the increasing trend of $\spt$ fluctuations with rapidity.

When both the energy density and net-baryon density normalizations are varied simultaneously by the same amount, their competing rapidity-dependent effects combine to determine the net fluctuation at a given rapidity. As illustrated in Fig. \ref{fig:2}, at two units of pseudorapidity away from midrapidity, the contribution from energy-density variation alone leads to a reduction of $v_0$ by approximately a factor of $\sim 3$ (noting that the vertical axis is shown on a logarithmic scale). In contrast, the variation of baryon density alone enhances $v_0$ by a factor of roughly $\sim 2$ over the same rapidity interval. Consequently, when both effects are present, their combined influence results in an overall decrease of $v_0$ by a factor of about $\sim 3/2$ at $|\eta|\simeq 2$ relative to midrapidity.

Most importantly, the results obtained with event-by-event fluctuating IC exhibit a rapidity dependence of $v_0$ that closely follows that observed for smooth IC. This agreement implies that the scaled observable $v_0(\eta)/v_0(\eta=0)$ is primarily sensitive to the longitudinal distributions of energy and baryon density, while being largely insensitive to transverse-plane fluctuations at different rapidities. This interpretation is closely aligned with the conclusions of Ref. \cite{Parida:2024ckk}, where the normalized quantity $v_0(p_T)/v_0$ was shown to be largely independent of collision centrality and initial-condition modeling details. These observations here reinforce the robustness of the midrapidity-scaled $\spt$ fluctuation observable as a sensitive probe of initial longitudinal profile in baryon-rich heavy-ion collisions.

As long as the transverse-plane–integrated rapidity distributions of energy density and net-baryon density are constrained to reproduce the measured rapidity distributions of charged particles and net protons, their longitudinal structures necessarily remain similar. Consequently, the resulting rapidity dependence of $v_0$, or equivalently of mean transverse momentum fluctuations, is expected to be model independent. Indeed, when I estimated the scaled observable $v_0(\eta)/v_0(\eta=0)$ using the results of Ref.~\cite{Liu:2025fbu}, where the calculations were performed with the NEOS-BQS equation of state, the magnitude of the $\eta$ dependence of $v_0(\eta)/v_0(\eta=0)$ is found to be almost close to that obtained in the present study. This agreement further supports the expectation that the scaled quantity is largely model independent.

The rapidity dependence of $\spt$ fluctuation can be measured in experiment. However, such measurements are affected by non-flow. Although self-correlation removal techniques can be applied, residual non-flow contributions may still persist \cite{STAR:2003cbv,CERES:2003sap,PHENIX:2003ccl,NA49:2003hxt,Heckel:2011mtu,STAR:2019dow,Chatterjee:2017mhc}. For this reason, observables that quantify $\spt$ correlations between two pseudorapidity windows separated by a finite rapidity gap provide a more robust experimental handle. In this context, the observables $R_{p_T}$ and the more strongly non-flow–suppressed three-bin correlator $r_{p_T}$ are particularly useful. These observables directly probe the degree to which transverse collective dynamics at different rapidities remain correlated.

\begin{figure}[th!]
    \includegraphics[width=\linewidth]{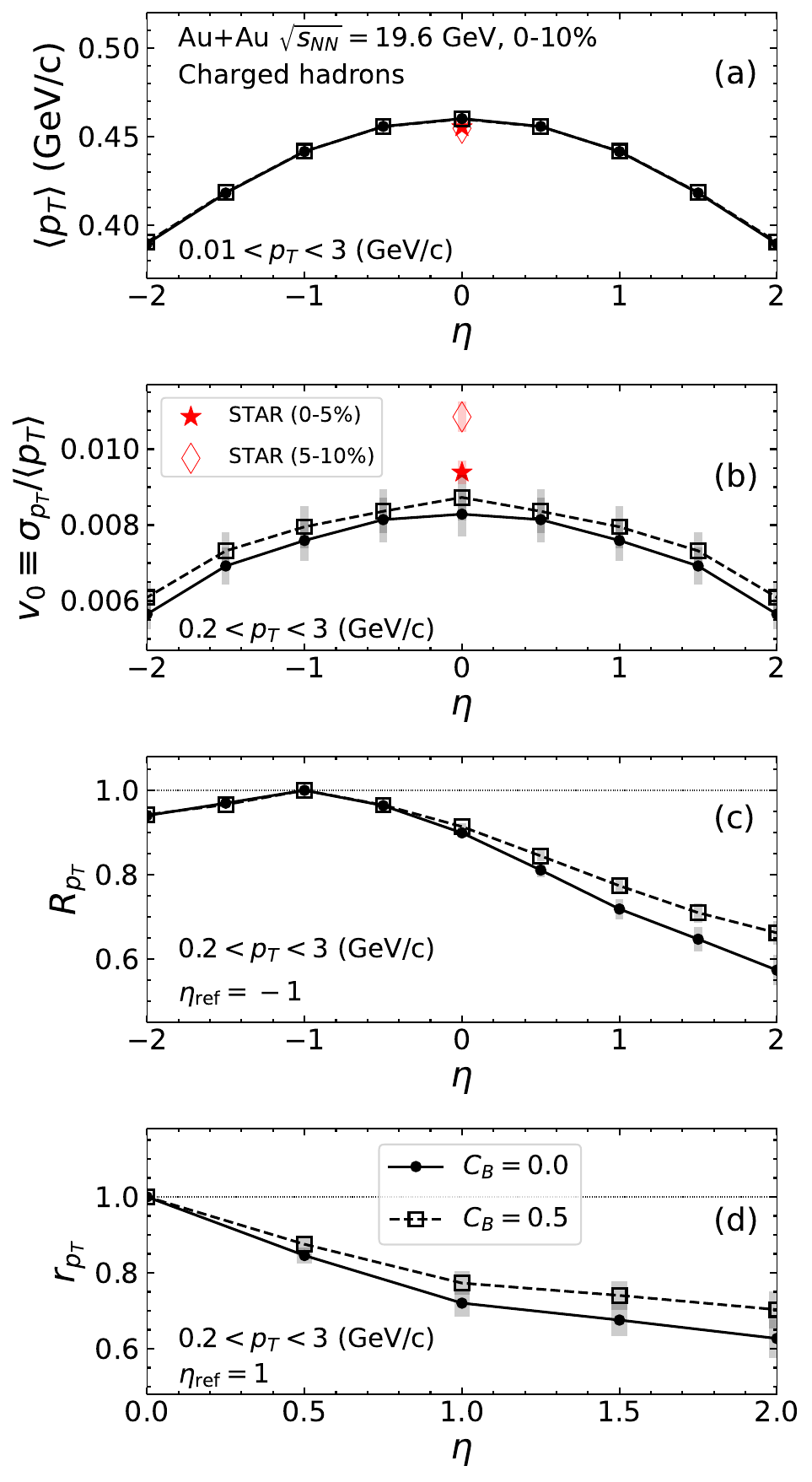}
    \caption{Pseudorapidity dependence of (a) $\langle p_T\rangle$, (b) $v_0 \equiv \sigma_{p_T}/\langle p_T\rangle$, (c) $R_{p_T}$, and (d) $r_{p_T}$ for charged hadrons in 0–10\% Au+Au collisions at $\sqrt{s_{NN}} = 19.6$ GeV. Model results with vanishing ($C_B = 0.0$) and finite ($C_B = 0.5$) baryon diffusion are shown for comparison. STAR midrapidity measurements are plotted where available \cite{STAR:2017sal,STAR:2019dow}.} 
        \label{fig:3}
\end{figure}

Recent studies, such as Ref.~\cite{Liu:2025fbu}, have demonstrated that the observables $R_{p_T}$ and $r_{p_T}$ are sensitive to the equation of state of the medium, while exhibiting only a weak dependence on shear and bulk viscosities. This makes them promising tools for constraining the speed of sound in baryon-rich matter. However, an important transport property has not yet been explored in this context, namely the baryon diffusion coefficient. The results presented in Fig.~\ref{fig:3} address this gap by investigating how baryon diffusion influences the rapidity dependence of mean-$p_T$ correlations.

It has been shown in Ref.~\cite{Denicol:2018wdp} that reproducing the experimentally measured rapidity distribution of net protons in the final state requires a modified rapidity profile of baryon deposition in the initial state when baryon diffusion is present. In particular, a nonzero baryon diffusion coefficient necessitates positioning the initial baryon-density peaks farther away from midrapidity. During the subsequent hydrodynamic evolution, additional diffusion-driven thermodynamic force transport more baryon toward midrapidity, thereby enabling a consistent description of the observed net-proton data.

Given that $\spt$ decorrelation observables are intrinsically sensitive to the longitudinal distribution of net-baryon density, it is therefore natural to expect that baryon diffusion can influence these observables. To investigate this effect, I explicitly study the impact of baryon diffusion. The results are presented in Fig.~\ref{fig:3}, where calculations with vanishing baryon diffusion are compared to those with nonzero baryon diffusion, corresponding to $C_B = 0.5$. It should be emphasized that, in the $C_B = 0.5$ case, the parameters of the initial baryon-density profile are appropriately tuned to reproduce the measured net-proton rapidity distribution.

In panel (a) of Fig.~\ref{fig:3}, I present the pseudorapidity dependence of the $\langle p_T \rangle$ for charged hadrons. For comparison, midrapidity measurements from the STAR Collaboration are also shown \cite{STAR:2017sal}. While STAR does not provide a direct measurement of the charged-hadron $\mpt$ at midrapidity, the corresponding results for identified pions, kaons, protons, and antiprotons are available. Using these data, I estimate the experimental charged-hadron $\mpt$, which is then compared with the model calculations. The model results correspond to the 0–10\% centrality class, whereas the experimental data are available for 0–5\% and 5–10\% centralities; both are therefore plotted for reference. In Fig.~\ref{fig:3}(b), I compare the STAR measurements of the relative $\spt$ fluctuation $\sigma_{p_T}/\langle p_T \rangle$ with the model results \cite{STAR:2019dow}. I observe that the model underestimates the magnitude of the experimental fluctuations, at midrapidity. This suggests that the present setup does not incorporate sufficient initial-state fluctuations. Introducing additional subnucleonic structure in the energy deposition, could enhance the agreement with the data. Notably, I find that the baryon diffusion coefficient $C_B$ has a negligible impact on both the mean transverse momentum and its fluctuation $v_0$, as seen in Fig.~\ref{fig:3}(a) and (b).

In Fig.~\ref{fig:3}(c) and (d), I show the rapidity dependece of $R_{p_T}$ and $r_{p_T}$, respectively, using $\eta_\text{ref} = -1$ and $\eta_\text{ref} = 1$ as the reference rapidity. To increase statistics, I symmetrize the expressions for $R_{p_T}(\eta; \eta_{\rm ref })$ and $r_{p_T}(\eta; \eta_{\rm ref})$ given in Eqs.~\ref{def:Rpt} and \ref{def:rpt} in the forward and backward rapidity directions. I observe the dependence of $R_{p_T}$ and $r_{p_T}$ on baryon diffusion is very weak and substantially smaller than the sensitivity to the equation of state reported in Ref.~\cite{Liu:2025fbu}. This observation indicates that these observables are robust against variations in baryon diffusion and therefore provide reliable probes for constraining the speed of sound in baryon-rich matter.

\begin{figure*}
 \begin{center}
  \includegraphics[scale=0.5]{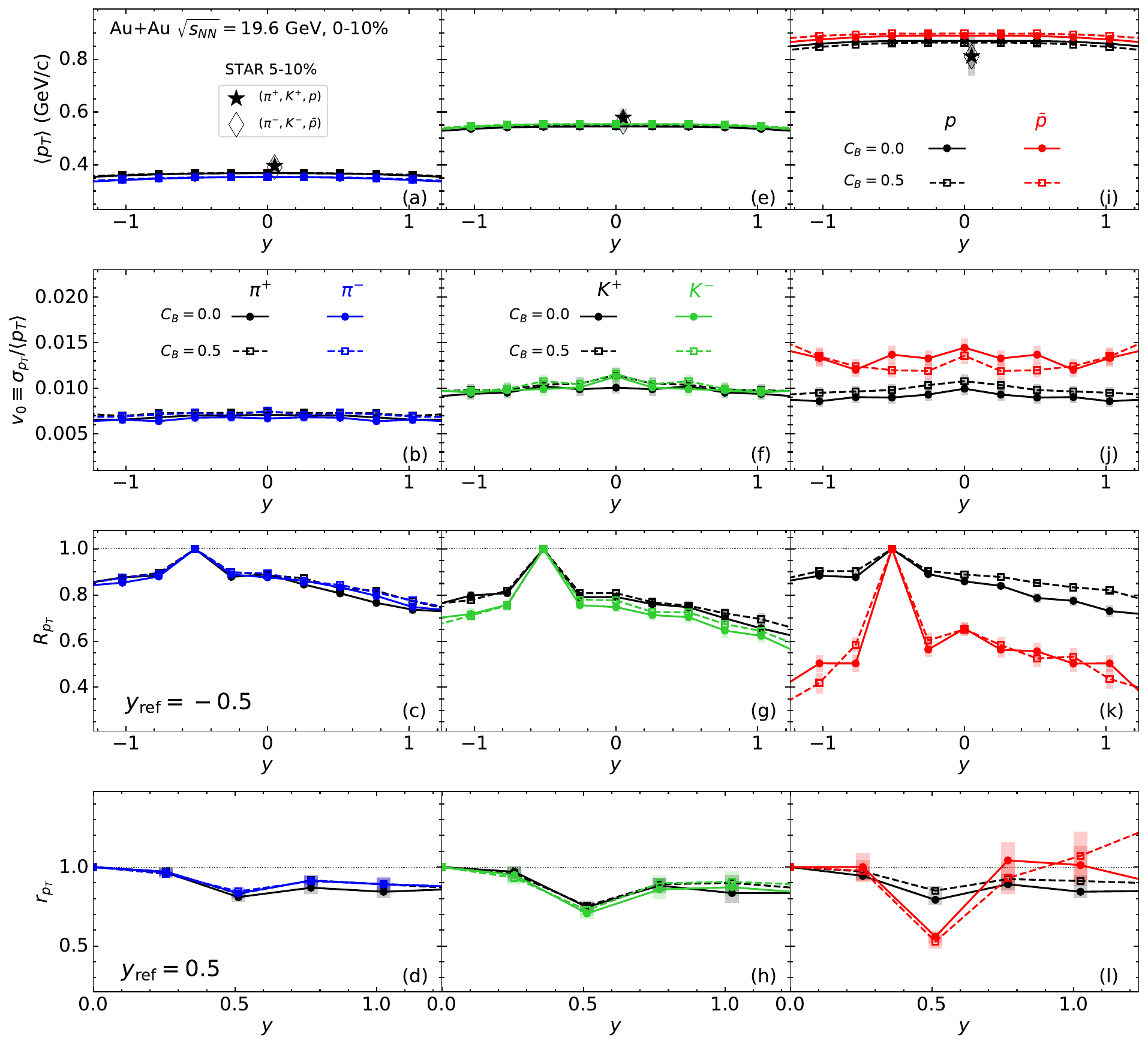}
 \caption{(Color Online) Rapidity dependence of observables related to mean transverse momentum correlations and fluctuations for identified hadrons in 0–10\% central Au+Au collisions at $\sqrt{s_{NN}} = 19.6$ GeV. Panels (a)–(d) show $\langle p_T\rangle$, $v_0=\sigma_{p_T}/\langle p_T\rangle$, $R_{p_T}$, and $r_{p_T}$ for $\pi^\pm$; panels (e)–(h) for $K^\pm$; and panels (i)–(l) for protons and antiprotons. Positively charged particles are shown with black lines and negatively charged particles with colored lines. Results with vanishing baryon diffusion ($C_B=0.0$, solid lines) are compared to those with finite baryon diffusion ($C_B=0.5$, dashed lines).  STAR data of $\mpt$ at 5-10\% centrality are plotted for comparison.}
 \label{fig:4}
 \end{center}
\end{figure*}

At lower collision energies, experimental measurements observed noticable splitting between baryons and antibaryons in both yield and anisotropic flow observables, originating due to the presence of a finite net-baryon density in the medium \cite{STAR:2017sal,STAR:2013ayu,STAR:2014clz}. Such splittings provide important insights into the initial-state baryon stopping and to the dependence of transport coefficients, such as shear and bulk viscosity, on the baryon chemical potential \cite{Shen:2020jwv,Parida:2023ldu,Parida:2022ppj,Bozek:2022svy,Du:2022yok}. The influence of finite baryon density on radial flow has also been investigated through studies of the $v_0(p_T)$ splitting between protons and antiprotons at RHIC BES energies, where model calculations predict a noticeable separation \cite{Jahan:2025cbp}. Given that the observables explored in this work are related to radial flow correlations and fluctuations across different rapidities, it is therefore natural to expect analogous baryon–antibaryon splittings in these quantities as well. This makes the study of $\spt$ fluctuations and decorrelations of identified particle particularly relevant, as they could provide a sensitive probe of the interplay between finite baryon density, collective dynamics, and longitudinal structure in heavy-ion collisions.

Fig.~\ref{fig:4} presents the rapidity dependence of $\spt$ decorrelation related observables for identified hadrons in 0–10\% Au+Au collisions at $\sqrt{s_{NN}}=19.6$ GeV, illustrating the effect of baryon diffusion. The left, middle, and right columns correspond to pions ($\pi^\pm$), kaons ($K^\pm$), and protons/antiprotons ($p$, $\bar p$), respectively. The top row shows the $\langle p_T\rangle$, the second row shows the relative fluctuation $v_0=\sigma_{p_T}/\langle p_T\rangle$, while the third and fourth rows display $R_{p_T}$ and $r_{p_T}$. I consider $\pi^\pm$,$K^\pm$ of $0.2<p_T<2$ GeV/c and $p,\bar{p}$ of $0.4<p_T<2$ GeV/c.

A clear mass dependence is observed in the $\mpt$ and relative fluctuation observable $v_0$ reflecting the expected mass ordering of radial flow \cite{Parida:2024ckk}; available STAR measurements of $\mpt$ are also shown for comparison \cite{STAR:2017sal}. Notably, however, a discernible splitting between protons and antiprotons emerges in $\sigma_{p_T}$. This splitting becomes more pronounced at larger rapidities, where the net-baryon density is higher, indicating that finite baryon density plays an increasingly important role in shaping transverse momentum fluctuations away from midrapidity.

The observable $R_{p_T}$ and $r_{p_T}$ is evaluated using the reference $y_{\mathrm{ref}} = -0.5$ and $y_{\mathrm{ref}} = 0.5$ respectively.  In Fig. \ref{fig:4}, a clear and pronounced splitting between protons and antiprotons is observed in $R_{p_T}$. From the definitions of $r_{p_T}$ and $R_{p_T}$, one can see that $r_{p_T}$ is a ratio of covariances, whereas $R_{p_T}$ contains in its numerator the covariance of $\spt$ between two rapidity bins and in its denominator the variance of $\spt$ in those bins. As discussed earlier, I observe a noticeable splitting in the variance of $\spt$ between protons and antiprotons. Therefore, in order to understand the origin of the splitting in $R_{p_T}$, it is necessary to disentangle the relative contributions from the covariance and the variance of $\spt$.

To this end, I calculated separately the covariance term appearing in the numerator of $R_{p_T}$ for protons and antiprotons, and quantified their difference by taking it's ratio to the sum of the covariances of protons and antiprotons. I find that the splitting in the covariance is around 10\%, whereas the total splitting observed in $R_{p_T}$ is approximately 30\%. This clearly indicates that the dominant contribution to the proton–antiproton splitting in $R_{p_T}$ originates from the difference in the variance of $\spt$, or equivalently in $\sigma_{p_T}$, rather than from the covariance term.
This observation suggests that, experimentally, it would be valuable to measure separately both the covariance of $\spt$ between two rapidity windows and the variance of $\spt$ within each window. In this context, the three-bin correlator $r_{p_T}$, which is constructed purely from covariances, provides a cleaner observable for studying baryon–antibaryon splitting in mean-$p_T$ decorrelation. However, due to limited statistical precision in my calculation, no definitive conclusion can be drawn regarding a possible proton–antiproton splitting in $r_{p_T}$, which probes transverse flow correlations between forward and backward rapidity bins placed symmetrically about midrapidity.

\begin{figure}[th!]
    \includegraphics[width=\linewidth]{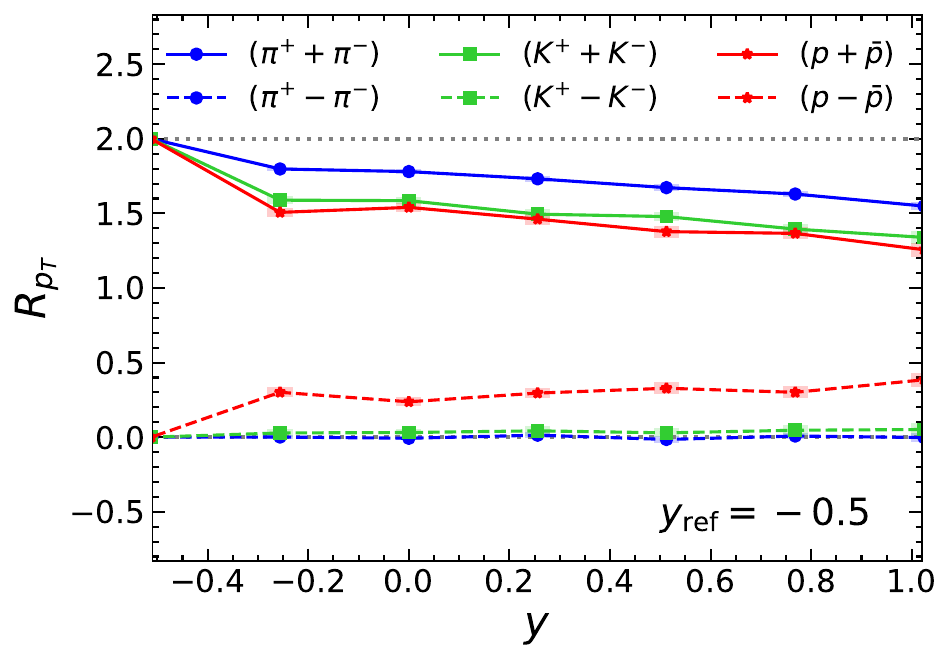}
    \caption{(Color Online) Rapidity dependence of $R_{p_T}$ for different identified-particle combinations in 0–10\% Au+Au collisions at $\sqrt{s_{NN}} = 19.6$ GeV. Results are shown for pion, kaon, and proton. Solid lines represents combinations with zero conserved quantum numbers, including $(\pi^+ + \pi^-)$, $(K^+ + K^-)$, and $(p + \bar p)$, where dashed lines represent combinations of nonzero conserved quantum numbers $(\pi^+ - \pi^+)$, $(K^+ - K^+)$, and $(p - \bar{p})$. The comparison highlights mass-dependent decorrelation as well as the splitting induced by finite conserved charges, particularly the baryon-density–driven separation between protons and antiprotons.} 
    \label{fig:5}
\end{figure}

In Fig. \ref{fig:4}(c), (g) and (k), the behavior of $R_{p_T}$ for identified particles reveals interesting features. While kaons exhibit a stronger decorrelation than pions, protons show a modest decorrelation with rapidity, whereas antiprotons display a significantly stronger decorrelation. To further elucidate this behavior, I examine combinations of particle species with zero net conserved quantum number \cite{Parida:2025ddt}, namely $(\pi^+ + \pi^-)$, $(K^+ + K^-)$, and $(p + \bar p)$, as shown in Fig.~\ref{fig:5}. The corresponding $R_{p_T}$ results exhibit a clear mass ordering, consistent with expectations from hydrodynamic evolution in the absence of net conserved charges.

I further construct particle combinations carrying nonzero conserved quantum numbers in order to isolate the role of conserved charges in $\spt$ decorrelation. The $(\pi^+-\pi^-)$ combination exhibits a vanishing $R_{p_T}$ splitting, indicating the absence of any significant effect associated with electric charge alone. In contrast, both the $(K^+-K^-)$ and $(p-\bar p)$ combinations show a finite splitting. In the calculations, I employ the NEOS-BQS equation of state \cite{Monnai:2019hkn}, which enforces strangeness neutrality and consequently introduces a nonzero strange chemical potential. This leads to a splitting between $K^+$ and $K^-$. Similarly, the presence of a finite baryon chemical potential results in a splitting between protons and antiprotons.

The observed splitting in $\sigma_{p_T}$ and $R_{p_T}$ between protons and antiprotons arises directly from the finite net-baryon density of the system; in the absence of baryon density, no such splitting would be expected. Previous measurements of the $\mpt$ for protons and antiprotons at RHIC Beam Energy Scan energies by STAR have shown prior indications of such baryon-antibaryon splitting in radial flow \cite{STAR:2017sal}, although large experimental uncertainties prevented firm conclusions. In contrast, the splitting observed here in $\spt$ fluctuations and their rapidity decorrelation reveals a large difference in the radial flow of protons and antiprotons in a baryon-rich medium. This suggests that measurements of both $\sigma_{p_T}$ and $R_{p_T}$ for identified hadrons could serve as a sensitive experimental probe of the initial baryon density distribution and the subsequent transverse collective dynamics of conserved charges.

In experimental measurements at lower \sNN, the yield of antiprotons is relatively small, making a direct determination of their rapidity decorrelation statistically challenging. However, as demonstrated in Fig. \ref{fig:5}, the $R_{p_T}$ should show a mass dependent behavior in the absence of conserved charges. This mass dependece is already predicted here quantitatively by the hydrodynamic calculation. Any additional deviation beyond this trivial mass dependence, particularly in the proton sector, can therefore be attributed to the longitudinal baryon density profile. Consequently, measuring the rapidity decorrelation of pions and protons alone is sufficient to place meaningful constraints on the underlying energy and baryon density fluctuations in a baryon-rich fireball.

\section{Summary}
\label{sec:summary}
In this work, I performed a systematic study of event-by-event mean transverse momentum $\spt$ fluctuations and their rapidity decorrelation in 0–10\% central Au+Au collisions at $\sqrt{s_{NN}} = 19.6$ GeV. The analysis focuses on the role of finite net-baryon density, baryon diffusion, and identified-particle dependence.
Using a hybrid hydrodynamics + hadronic transport framework with conserved baryon current, I investigated how the longitudinal distributions of energy and baryon density influence transverse radial flow fluctuations and correlations across rapidity.

I first demonstrated that the rapidity dependence of the scaled $\spt$ fluctuation $\frac{\sigma_{p_T}}{\mpt}$($ = v_0$) is governed by a competition between energy-density and baryon-density fluctuations. Variations in energy density lead to a decreasing $v_0$ with increasing rapidity, whereas baryon-density fluctuations generate an opposite trend due to the growth of net-baryon density away from midrapidity. When both effects are present, their interplay determines the observed rapidity dependence. Importantly, I showed that the scaled observable $v_0(\eta)/v_0(\eta=0)$ exhibits a similar rapidity dependence for both smooth and fluctuating IC. This indicates that this quantity is largely independent of transverse fluctuations and is primarily controlled by the longitudinal profile of energy and baryon density, making it a robust observable for constraining the equation of state in baryon-rich matter.

I then studied experimentally accessible $\spt$ correlation observables, namely $R_{p_T}$ and the three-bin correlator $r_{p_T}$, which are expected to be less affected by non-flow contributions. I found that the rapidity dependence of both observables exhibit only a weak sensitivity to baryon diffusion, while previous studies have demonstrated a significant dependence on the equation of state. This establishes $R_{p_T}$ and $r_{p_T}$ as robust probes of the speed of sound in baryon-dense matter, with minimal contamination from baryon diffusion effects.

Extending the analysis to identified particles, I observed the expected mass ordering in $\langle p_T\rangle$ and $v_0$. More importantly, a clear baryon–antibaryon splitting emerges in both $v_0$ and $R_{p_T}$ for protons and antiprotons. This splitting reflects differences in radial flow dynamics between baryons and antibaryons in a baryon-rich medium. By constructing particle combinations with zero and nonzero conserved quantum numbers, I further isolated the role of conserved charges \cite{Parida:2025ddt}. I showed that combinations such as $(\pi^+ + \pi^-)$, $(K^+ + K^-)$, and $(p + \bar p)$ exhibit only a trivial mass-driven decorrelation, while combinations with nonzero conserved charges, such as $(K^+ - K^-)$ and $(p - \bar p)$, display a finite splitting arising from nonzero strange and baryon chemical potentials, respectively.

Finally, I note several limitations of the present study. The model does not include event-by-event fluctuations in initial longitudinal profile of baryon or energy deposition \cite{Broniowski:2016ybk,Shen:2017bsr}, nor does it implement a dynamical initialization of hydrodynamics \cite{Shen:2023awv,Du:2018mpf}; instead, hydrodynamic evolution is started at a fixed proper time. In addition, pre-equilibrium dynamics and possible deviations from initial Bjorken longitudinal flow are not included \cite{Du:2025bhb,Ryu:2021lnx}. These effects may quantitatively modify the predicted magnitude of mean-$p_T$ fluctuation and it's rapidity decorrelation. Nevertheless, I expect the qualitative physics conclusions presented here to remain useful.

Overall, the results show that the rapidity dependence of mean transverse momentum fluctuations and their decorrelation, for both charged and identified hadron, provides a sensitive and experimentally feasible probe of the three-dimensional initial-state structure, baryon stopping, equation of state, and collective dynamics in heavy-ion collisions at finite baryon density.

\section{Data Availability}
The numerical data corresponding to the figures presented in this paper are openly available in Ref.~\cite{Parida:2026data}.

\section{Acknowledgement}
This work is supported by the AGH University of Krakow and by the Polish National Science Centre grant: 2023/51/B/ST2/01625. In preparation of this publication I used the resources of the Centre for Computation and Computer Modelling of the Jan Kochanowski University in Kielce. I am thankful to Piotr Bożek for carefully reading the manuscript and providing valuable comments and suggestions. I would also like to thank Prabhupada Dixit for helpful comment on oversampling technique and the reduction of statistical fluctuation. In addition, I am thankful to Jean-Yves Ollitrault for insightful discussions on mean-$p_T$ fluctuations in other related works, which greatly benefited this study.

\appendix

\section{Effect of the Baryon Deposition Parameter $\omega$}
\label{app:omega_zero}

In this appendix, I study the effect of different schemes for the transverse deposition of the initial baryon density on the qualitative features of the rapidity dependence of mean $p_T$ fluctuations. In the setup used in Sec.~\ref{sec:results} with $C_B=0$, I employ $\omega=0.13$, which incorporates a contribution from binary collisions in the transverse baryon deposition (see Eq.~\eqref{eq:nb3d}, where $\omega$ controls the relative contribution and thereby the transverse structure of the initial net-baryon density).

However, as discussed in Sec.~\ref{sec:model}, a purely participant-driven transverse baryon deposition is expected to capture the essential physics and not alter the qualitative behavior of the observable studied in this work. To verify this expectation, I consider the case $\omega=0$, corresponding to purely participant-driven baryon deposition. I repeat the same calculation as performed for Fig.~\ref{fig:2}, using smooth IC with $C_B=0$, and compare the results with the case $\omega=0.13$ used in the main text.

\begin{figure}[th!]
    \includegraphics[width=\linewidth]{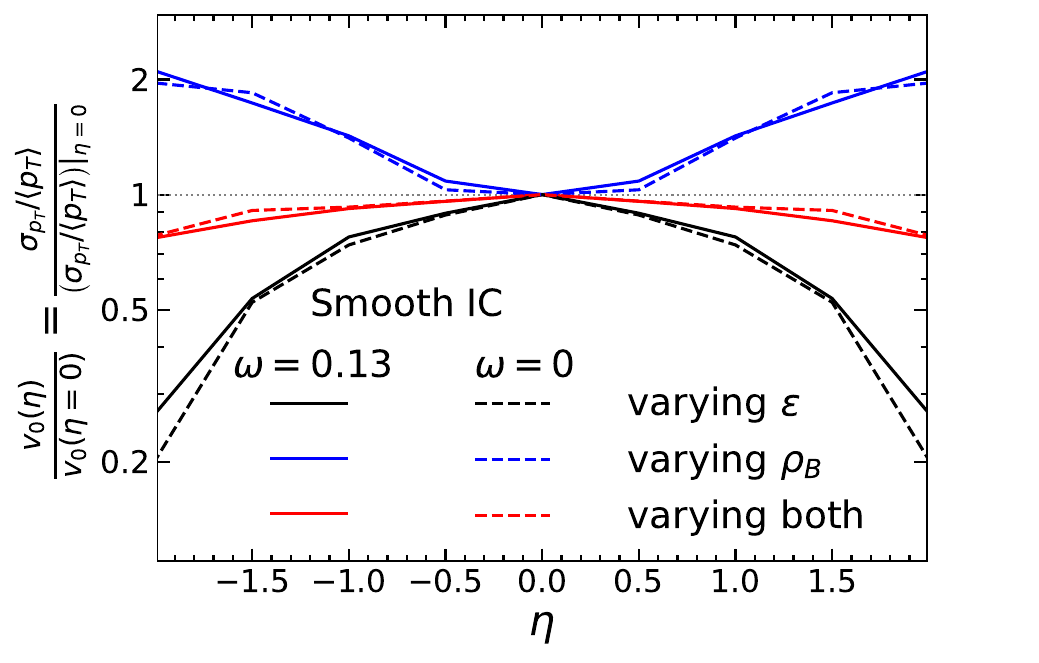}
    \caption{(Color Online) Rapidity dependence of the scaled observable $v_0(\eta)/v_0(\eta=0)$ calculated using smooth IC with $C_B=0$. Results are shown for the case $\omega=0.13$ (solid lines), which includes a binary collision contribution to baryon deposition, and for $\omega=0$ (dashed lines), corresponding to purely participant-driven baryon deposition. The different colors represent independent variations of energy density ($\epsilon$, black), baryon density ($\rho_B$, blue), and simultaneous variation of both (red), following the same procedure as in Fig. \ref{fig:2}.} 
    \label{fig:omega_zero}
\end{figure}

In Fig.~\ref{fig:omega_zero}, I show the rapidity dependence of the scaled observable $v_0(\eta)/v_0(\eta=0)$ for both $\omega=0.13$ (solid lines) and $\omega=0$ (dashed lines). As in Fig.~\ref{fig:2}, I separately vary the energy density ($\epsilon$), baryon density ($\rho_B$), and both simultaneously.

I find that the qualitative features of $v_0(\eta)/v_0(\eta=0)$, including its rapidity dependence and the relative ordering between variations of energy density and net-baryon density, remain unchanged when going from $\omega=0.13$ to $\omega=0$. This demonstrates that the conclusions of this work are robust with respect to the choice of transverse baryon deposition scheme and strengthens the claim that the observed behavior is largely independent of the details of the initial-state modeling.

\begin{thebibliography}{99}
\bibitem{Ollitrault:2007du}
J.~Y.~Ollitrault,
Eur. J. Phys. \textbf{29} (2008), 275-302
doi:10.1088/0143-0807/29/2/010
[arXiv:0708.2433 [nucl-th]].

\bibitem{Ollitrault:1992bk}
J.~Y.~Ollitrault,
Phys. Rev. D \textbf{46} (1992), 229-245
doi:10.1103/PhysRevD.46.229


\bibitem{Bozek:2014cva}
P.~Bozek, W.~Broniowski, E.~Ruiz Arriola and M.~Rybczynski,
Phys. Rev. C \textbf{90} (2014) no.6, 064902
doi:10.1103/PhysRevC.90.064902
[arXiv:1410.7434 [nucl-th]].





\bibitem{Song:2010mg}
H.~Song, S.~A.~Bass, U.~Heinz, T.~Hirano and C.~Shen,
Phys. Rev. Lett. \textbf{106} (2011), 192301
[erratum: Phys. Rev. Lett. \textbf{109} (2012), 139904]
doi:10.1103/PhysRevLett.106.192301
[arXiv:1011.2783 [nucl-th]].

\bibitem{Bozek:2009dw}
P.~Bozek,
Phys. Rev. C \textbf{81} (2010), 034909
doi:10.1103/PhysRevC.81.034909
[arXiv:0911.2397 [nucl-th]].

\bibitem{Bozek:2011ua}
P.~Bozek,
Phys. Rev. C \textbf{85} (2012), 034901
doi:10.1103/PhysRevC.85.034901
[arXiv:1110.6742 [nucl-th]].

\bibitem{Karpenko:2015xea}
I.~A.~Karpenko, P.~Huovinen, H.~Petersen and M.~Bleicher,
Phys. Rev. C \textbf{91} (2015) no.6, 064901
doi:10.1103/PhysRevC.91.064901
[arXiv:1502.01978 [nucl-th]].

\bibitem{Shen:2015msa}
C.~Shen and U.~Heinz,
Nucl. Phys. News \textbf{25} (2015) no.2, 6-11
doi:10.1080/10619127.2015.1006502
[arXiv:1507.01558 [nucl-th]].

\bibitem{Ryu:2015vwa}
S.~Ryu, J.~F.~Paquet, C.~Shen, G.~S.~Denicol, B.~Schenke, S.~Jeon and C.~Gale,
Phys. Rev. Lett. \textbf{115} (2015) no.13, 132301
doi:10.1103/PhysRevLett.115.132301
[arXiv:1502.01675 [nucl-th]].

\bibitem{Shen:2020jwv}
C.~Shen and S.~Alzhrani,
Phys. Rev. C \textbf{102} (2020) no.1, 014909
doi:10.1103/PhysRevC.102.014909
[arXiv:2003.05852 [nucl-th]].

\bibitem{OmanaKuttan:2022aml}
M.~Omana Kuttan, J.~Steinheimer, K.~Zhou and H.~Stoecker,
Phys. Rev. Lett. \textbf{131} (2023) no.20, 202303
doi:10.1103/PhysRevLett.131.202303
[arXiv:2211.11670 [hep-ph]].

\bibitem{Gong:2024lhq}
J.~Gong, H.~Roch and C.~Shen,
Phys. Rev. C \textbf{111} (2025) no.4, 044912
doi:10.1103/PhysRevC.111.044912
[arXiv:2410.22160 [nucl-th]].

\bibitem{STAR:2000ekf}
K.~H.~Ackermann \textit{et al.} [STAR],
Phys. Rev. Lett. \textbf{86} (2001), 402-407
doi:10.1103/PhysRevLett.86.402
[arXiv:nucl-ex/0009011 [nucl-ex]].

\bibitem{ALICE:2011ab}
K.~Aamodt \textit{et al.} [ALICE],
Phys. Rev. Lett. \textbf{107} (2011), 032301
doi:10.1103/PhysRevLett.107.032301
[arXiv:1105.3865 [nucl-ex]].

\bibitem{STAR:2015rxv}
L.~Adamczyk \textit{et al.} [STAR],
Phys. Rev. C \textbf{93} (2016) no.1, 014907
doi:10.1103/PhysRevC.93.014907
[arXiv:1509.08397 [nucl-ex]].


\bibitem{PHENIX:2003qra}
S.~S.~Adler \textit{et al.} [PHENIX],
Phys. Rev. Lett. \textbf{91} (2003), 182301
doi:10.1103/PhysRevLett.91.182301
[arXiv:nucl-ex/0305013 [nucl-ex]].


\bibitem{Bernhard:2016tnd}
J.~E.~Bernhard, J.~S.~Moreland, S.~A.~Bass, J.~Liu and U.~Heinz,
Phys. Rev. C \textbf{94} (2016) no.2, 024907
doi:10.1103/PhysRevC.94.024907
[arXiv:1605.03954 [nucl-th]].


\bibitem{Bernhard:2019bmu}
J.~E.~Bernhard, J.~S.~Moreland and S.~A.~Bass,
Nature Phys. \textbf{15} (2019) no.11, 1113-1117
doi:10.1038/s41567-019-0611-8

\bibitem{Nijs:2020ors}
G.~Nijs, W.~van der Schee, U.~G{\"u}rsoy and R.~Snellings,
Phys. Rev. Lett. \textbf{126} (2021) no.20, 202301
doi:10.1103/PhysRevLett.126.202301
[arXiv:2010.15130 [nucl-th]].

\bibitem{JETSCAPE:2020mzn}
D.~Everett \textit{et al.} [JETSCAPE],
Phys. Rev. C \textbf{103} (2021) no.5, 054904
doi:10.1103/PhysRevC.103.054904
[arXiv:2011.01430 [hep-ph]].

\bibitem{Jahan:2024wpj}
S.~A.~Jahan, H.~Roch and C.~Shen,
Phys. Rev. C \textbf{110} (2024) no.5, 054905
doi:10.1103/PhysRevC.110.054905
[arXiv:2408.00537 [nucl-th]].





\bibitem{Schenke:2020uqq}
B.~Schenke, C.~Shen and D.~Teaney,
Phys. Rev. C \textbf{102} (2020) no.3, 034905
[arXiv:2004.00690 [nucl-th]].

\bibitem{Samanta:2025yrj}
R.~Samanta and T.~Parida,
J. Subatomic Part. Cosmol. \textbf{4} (2025), 100100
doi:10.1016/j.jspc.2025.100100
[arXiv:2505.02706 [nucl-th]].

\bibitem{Parida:2024ckk}
T.~Parida, R.~Samanta and J.~Y.~Ollitrault,
Phys. Lett. B \textbf{857} (2024), 138985
doi:10.1016/j.physletb.2024.138985
[arXiv:2407.17313 [nucl-th]].

\bibitem{Samanta:2023amp}
R.~Samanta, S.~Bhatta, J.~Jia, M.~Luzum and J.~Y.~Ollitrault,
Phys. Rev. C \textbf{109} (2024) no.5, L051902
doi:10.1103/PhysRevC.109.L051902
[arXiv:2303.15323 [nucl-th]].

\bibitem{Jia:2025rab}
J.~Jia,
[arXiv:2507.14399 [nucl-th]].

\bibitem{Bhatta:2025oyp}
S.~Bhatta, A.~Dimri and J.~Jia,
[arXiv:2504.20008 [nucl-th]].

\bibitem{Zhou:2025bwu}
F.~Zhou, G.~Giacalone and J.~Y.~Ollitrault,
[arXiv:2511.04605 [nucl-th]].

\bibitem{Giacalone:2020lbm}
G.~Giacalone, F.~G.~Gardim, J.~Noronha-Hostler and J.~Y.~Ollitrault,
Phys. Rev. C \textbf{103} (2021) no.2, 024910
doi:10.1103/PhysRevC.103.024910
[arXiv:2004.09799 [nucl-th]].

\bibitem{Gardim:2019brr}
F.~G.~Gardim, G.~Giacalone and J.~Y.~Ollitrault,
Phys. Lett. B \textbf{809} (2020), 135749
doi:10.1016/j.physletb.2020.135749
[arXiv:1909.11609 [nucl-th]].

\bibitem{Saha:2025nyu}
S.~Saha, R.~Singh and B.~Mohanty,
Phys. Rev. C \textbf{112} (2025) no.2, 024902
doi:10.1103/83zq-kdjg
[arXiv:2505.19697 [nucl-ex]].

\bibitem{Broniowski:2009fm}
W.~Broniowski, M.~Chojnacki and L.~Obara,
Phys. Rev. C \textbf{80} (2009), 051902
doi:10.1103/PhysRevC.80.051902
[arXiv:0907.3216 [nucl-th]].


\bibitem{Bozek:2017elk}
P.~Bo{\.z}ek and W.~Broniowski,
Phys. Rev. C \textbf{96} (2017) no.1, 014904
doi:10.1103/PhysRevC.96.014904
[arXiv:1701.09105 [nucl-th]].

\bibitem{Mu:2025gtr}
Y.~S.~Mu, J.~A.~Sun, L.~Yan and X.~G.~Huang,
Phys. Rev. Lett. \textbf{135} (2025) no.16, 162301
doi:10.1103/skhj-cj9p
[arXiv:2501.02777 [nucl-th]].

\bibitem{Gardim:2019xjs}
F.~G.~Gardim, G.~Giacalone, M.~Luzum and J.~Y.~Ollitrault,
Nature Phys. \textbf{16} (2020) no.6, 615-619
doi:10.1038/s41567-020-0846-4
[arXiv:1908.09728 [nucl-th]].

\bibitem{Gardim:2019brr}
F.~G.~Gardim, G.~Giacalone and J.~Y.~Ollitrault,
Phys. Lett. B \textbf{809} (2020), 135749
doi:10.1016/j.physletb.2020.135749
[arXiv:1909.11609 [nucl-th]].

\bibitem{CMS:2023byu}
 [CMS],
CMS-PAS-HIN-23-003.

\bibitem{Parida:2024ckk}
T.~Parida, R.~Samanta and J.~Y.~Ollitrault,
Phys. Lett. B \textbf{857} (2024), 138985
doi:10.1016/j.physletb.2024.138985
[arXiv:2407.17313 [nucl-th]].

\bibitem{ALICE:2025iud}
S.~Acharya \textit{et al.} [ALICE],
[arXiv:2504.04796 [nucl-ex]].

\bibitem{ATLAS:2025ztg}
G.~Aad \textit{et al.} [ATLAS],
[arXiv:2503.24125 [nucl-ex]].

\bibitem{Du:2025hrz}
L.~Du and P.~M.~Jacobs,
[arXiv:2512.10265 [nucl-th]].


\bibitem{Du:2025dpu}
L.~Du,
[arXiv:2508.07184 [hep-ph]].

\bibitem{Ratti:2018ksb}
C.~Ratti,
Rept. Prog. Phys. \textbf{81} (2018) no.8, 084301
doi:10.1088/1361-6633/aabb97
[arXiv:1804.07810 [hep-lat]].

\bibitem{Monnai:2019hkn}
A.~Monnai, B.~Schenke and C.~Shen,
Phys. Rev. C \textbf{100} (2019) no.2, 024907
doi:10.1103/PhysRevC.100.024907
[arXiv:1902.05095 [nucl-th]].

\bibitem{Monnai:2024pvy}
A.~Monnai, G.~Pihan, B.~Schenke and C.~Shen,
Phys. Rev. C \textbf{110} (2024) no.4, 044905
doi:10.1103/PhysRevC.110.044905
[arXiv:2406.11610 [nucl-th]].

\bibitem{Noronha-Hostler:2019ayj}
J.~Noronha-Hostler, P.~Parotto, C.~Ratti and J.~M.~Stafford,
Phys. Rev. C \textbf{100} (2019) no.6, 064910
doi:10.1103/PhysRevC.100.064910
[arXiv:1902.06723 [hep-ph]].

\bibitem{Mondal:2021jxk}
S.~Mondal, S.~Mukherjee and P.~Hegde,
Phys. Rev. Lett. \textbf{128} (2022) no.2, 022001
doi:10.1103/PhysRevLett.128.022001
[arXiv:2106.03165 [hep-lat]].

\bibitem{Bjorken:1982qr}
J.~D.~Bjorken,
Phys. Rev. D \textbf{27} (1983), 140-151
doi:10.1103/PhysRevD.27.140

\bibitem{Dumitru:2008wn}
A.~Dumitru, F.~Gelis, L.~McLerran and R.~Venugopalan,
Nucl. Phys. A \textbf{810} (2008), 91-108
doi:10.1016/j.nuclphysa.2008.06.012
[arXiv:0804.3858 [hep-ph]].


\bibitem{Giacalone:2020ymy}
G.~Giacalone,
[arXiv:2101.00168 [nucl-th]].

\bibitem{Bhalerao:2019fzp}
R.~S.~Bhalerao, G.~Giacalone and J.~Y.~Ollitrault,
Phys. Rev. C \textbf{100} (2019) no.1, 014909
doi:10.1103/PhysRevC.100.014909
[arXiv:1904.10350 [nucl-th]].

\bibitem{Gelis:2019vzt}
F.~Gelis, G.~Giacalone, P.~Guerrero-Rodr{\'\i}guez, C.~Marquet and J.~Y.~Ollitrault,
[arXiv:1907.10948 [nucl-th]].


\bibitem{Samanta:2025fuj}
R.~Samanta,
[arXiv:2505.12961 [nucl-th]].

\bibitem{Song:2009gc}
H.~Song,
[arXiv:0908.3656 [nucl-th]].

\bibitem{Luzum:2009ue}
M.~Luzum,
[arXiv:0908.4100 [nucl-th]].

\bibitem{Noronha-Hostler:2013gga}
J.~Noronha-Hostler, G.~S.~Denicol, J.~Noronha, R.~P.~G.~Andrade and F.~Grassi,
Phys. Rev. C \textbf{88} (2013) no.4, 044916
doi:10.1103/PhysRevC.88.044916
[arXiv:1305.1981 [nucl-th]].

\bibitem{Bernhard:2018hnz}
J.~E.~Bernhard,
[arXiv:1804.06469 [nucl-th]].

\bibitem{Roy:2012jb}
V.~Roy, A.~K.~Chaudhuri and B.~Mohanty,
Phys. Rev. C \textbf{86} (2012), 014902
doi:10.1103/PhysRevC.86.014902
[arXiv:1204.2347 [nucl-th]].

\bibitem{Song:2011hk}
H.~Song, S.~A.~Bass, U.~Heinz, T.~Hirano and C.~Shen,
Phys. Rev. C \textbf{83} (2011), 054910
[erratum: Phys. Rev. C \textbf{86} (2012), 059903]
doi:10.1103/PhysRevC.83.054910
[arXiv:1101.4638 [nucl-th]].


\bibitem{Pang:2014pxa}
L.~G.~Pang, G.~Y.~Qin, V.~Roy, X.~N.~Wang and G.~L.~Ma,
Phys. Rev. C \textbf{91} (2015) no.4, 044904
doi:10.1103/PhysRevC.91.044904
[arXiv:1410.8690 [nucl-th]].

\bibitem{Pang:2015zrq}
L.~G.~Pang, H.~Petersen, G.~Y.~Qin, V.~Roy and X.~N.~Wang,
Eur. Phys. J. A \textbf{52} (2016) no.4, 97
doi:10.1140/epja/i2016-16097-x
[arXiv:1511.04131 [nucl-th]].


\bibitem{Nie:2022gbg}
M.~Nie, C.~Zhang, Z.~Chen, L.~Yi and J.~Jia,
Phys. Lett. B \textbf{845} (2023), 138177
doi:10.1016/j.physletb.2023.138177
[arXiv:2208.05416 [nucl-th]].


\bibitem{Chatterjee:2017mhc}
S.~Chatterjee and P.~Bozek,
Phys. Rev. C \textbf{96} (2017) no.1, 014906
doi:10.1103/PhysRevC.96.014906
[arXiv:1704.02777 [nucl-th]].

\bibitem{Bozek:2015bha}
P.~Bo{\.z}ek, W.~Broniowski and A.~Olszewski,
Phys. Rev. C \textbf{91} (2015), 054912
doi:10.1103/PhysRevC.91.054912
[arXiv:1503.07425 [nucl-th]].

\bibitem{Bozek:2015bna}
P.~Bozek and W.~Broniowski,
Phys. Lett. B \textbf{752} (2016), 206-211
doi:10.1016/j.physletb.2015.11.054
[arXiv:1506.02817 [nucl-th]].

\bibitem{ATLAS:2020sgl}
G.~Aad \textit{et al.} [ATLAS],
Phys. Rev. Lett. \textbf{126} (2021) no.12, 122301
doi:10.1103/PhysRevLett.126.122301
[arXiv:2001.04201 [nucl-ex]].



\bibitem{ATLAS:2017rij}
M.~Aaboud \textit{et al.} [ATLAS],
Eur. Phys. J. C \textbf{78} (2018) no.2, 142
doi:10.1140/epjc/s10052-018-5605-7
[arXiv:1709.02301 [nucl-ex]].

\bibitem{CMS:2015xmx}
V.~Khachatryan \textit{et al.} [CMS],
Phys. Rev. C \textbf{92} (2015) no.3, 034911
doi:10.1103/PhysRevC.92.034911
[arXiv:1503.01692 [nucl-ex]].

\bibitem{Jia:2014ysa}
J.~Jia and P.~Huo,
Phys. Rev. C \textbf{90} (2014) no.3, 034915
doi:10.1103/PhysRevC.90.034915
[arXiv:1403.6077 [nucl-th]].




\bibitem{Huo:2013qma}
P.~Huo, J.~Jia and S.~Mohapatra,
Phys. Rev. C \textbf{90} (2014) no.2, 024910
doi:10.1103/PhysRevC.90.024910
[arXiv:1311.7091 [nucl-ex]].

\bibitem{Xiao:2012uw}
K.~Xiao, F.~Liu and F.~Wang,
Phys. Rev. C \textbf{87} (2013) no.1, 011901
doi:10.1103/PhysRevC.87.011901
[arXiv:1208.1195 [nucl-th]].

\bibitem{Bzdak:2012tp}
A.~Bzdak and D.~Teaney,
Phys. Rev. C \textbf{87} (2013) no.2, 024906
doi:10.1103/PhysRevC.87.024906
[arXiv:1210.1965 [nucl-th]].

\bibitem{Bozek:2010vz}
P.~Bozek, W.~Broniowski and J.~Moreira,
Phys. Rev. C \textbf{83} (2011), 034911
doi:10.1103/PhysRevC.83.034911
[arXiv:1011.3354 [nucl-th]].



\bibitem{STAR:2014clz}
L.~Adamczyk \textit{et al.} [STAR],
Phys. Rev. Lett. \textbf{112} (2014) no.16, 162301
doi:10.1103/PhysRevLett.112.162301
[arXiv:1401.3043 [nucl-ex]].

\bibitem{STAR:2019vcp}
J.~Adam \textit{et al.} [STAR],
Phys. Rev. C \textbf{101} (2020) no.2, 024905
doi:10.1103/PhysRevC.101.024905
[arXiv:1908.03585 [nucl-ex]].

\bibitem{Nie:2020trj}
M.~Nie [STAR],
Nucl. Phys. A \textbf{1005} (2021), 121783
doi:10.1016/j.nuclphysa.2020.121783
[arXiv:2005.03252 [nucl-ex]].

\bibitem{Cimerman:2021gwf}
J.~Cimerman, I.~Karpenko, B.~Tom{\'a}{\v{s}}ik and B.~A.~Trzeciak,
Phys. Rev. C \textbf{104} (2021) no.1, 014904
doi:10.1103/PhysRevC.104.014904
[arXiv:2104.08022 [nucl-th]].


\bibitem{Back:2002wb}
B.~B.~Back, M.~D.~Baker, D.~S.~Barton, R.~R.~Betts, M.~Ballintijn, A.~A.~Bickley, R.~Bindel, A.~Budzanowski, W.~Busza and A.~Carroll, \textit{et al.}
Phys. Rev. Lett. \textbf{91} (2003), 052303
doi:10.1103/PhysRevLett.91.052303
[arXiv:nucl-ex/0210015 [nucl-ex]].

\bibitem{BRAHMS:2003wwg}
I.~G.~Bearden \textit{et al.} [BRAHMS],
Phys. Rev. Lett. \textbf{93} (2004), 102301
doi:10.1103/PhysRevLett.93.102301
[arXiv:nucl-ex/0312023 [nucl-ex]].

\bibitem{NA49:1998gaz}
H.~Appelshauser \textit{et al.} [NA49],
Phys. Rev. Lett. \textbf{82} (1999), 2471-2475
doi:10.1103/PhysRevLett.82.2471
[arXiv:nucl-ex/9810014 [nucl-ex]].

\bibitem{Liu:2025fbu}
L.~M.~Liu, J.~Chen, X.~G.~Huang, J.~Jia, C.~Shen and C.~Zhang,
[arXiv:2511.11094 [nucl-th]].

\bibitem{ATLAS:2022dov}
G.~Aad \textit{et al.} [ATLAS],
Phys. Rev. C \textbf{107} (2023) no.5, 054910
doi:10.1103/PhysRevC.107.054910
[arXiv:2205.00039 [nucl-ex]].

\bibitem{Schenke:2010nt}
B.~Schenke, S.~Jeon and C.~Gale,
Phys. Rev. C \textbf{82} (2010), 014903
doi:10.1103/PhysRevC.82.014903
[arXiv:1004.1408 [hep-ph]].

\bibitem{Schenke:2011bn}
B.~Schenke, S.~Jeon and C.~Gale,
Phys. Rev. C \textbf{85} (2012), 024901
doi:10.1103/PhysRevC.85.024901
[arXiv:1109.6289 [hep-ph]].

\bibitem{Denicol:2018wdp}
G.~S.~Denicol, C.~Gale, S.~Jeon, A.~Monnai, B.~Schenke and C.~Shen,
Phys. Rev. C \textbf{98} (2018) no.3, 034916
doi:10.1103/PhysRevC.98.034916
[arXiv:1804.10557 [nucl-th]].

\bibitem{Bass:1998ca}
S.~A.~Bass, M.~Belkacem, M.~Bleicher, M.~Brandstetter, L.~Bravina, C.~Ernst, L.~Gerland, M.~Hofmann, S.~Hofmann and J.~Konopka, \textit{et al.}
Prog. Part. Nucl. Phys. \textbf{41} (1998), 255-369
doi:10.1016/S0146-6410(98)00058-1
[arXiv:nucl-th/9803035 [nucl-th]].


\bibitem{Bleicher:1999xi}
M.~Bleicher, E.~Zabrodin, C.~Spieles, S.~A.~Bass, C.~Ernst, S.~Soff, L.~Bravina, M.~Belkacem, H.~Weber and H.~Stoecker, \textit{et al.}
J. Phys. G \textbf{25} (1999), 1859-1896
doi:10.1088/0954-3899/25/9/308
[arXiv:hep-ph/9909407 [hep-ph]].

\bibitem{Shen:2014vra}
C.~Shen, Z.~Qiu, H.~Song, J.~Bernhard, S.~Bass and U.~Heinz,
Comput. Phys. Commun. \textbf{199} (2016), 61-85
doi:10.1016/j.cpc.2015.08.039
[arXiv:1409.8164 [nucl-th]].

\bibitem{Schenke:2020mbo}
B.~Schenke, C.~Shen and P.~Tribedy,
Phys. Rev. C \textbf{102} (2020) no.4, 044905
doi:10.1103/PhysRevC.102.044905
[arXiv:2005.14682 [nucl-th]].

\bibitem{Bozek:2010bi}
P.~Bozek and I.~Wyskiel,
Phys. Rev. C \textbf{81} (2010), 054902
doi:10.1103/PhysRevC.81.054902
[arXiv:1002.4999 [nucl-th]].

\bibitem{Parida:2022lmt}
T.~Parida and S.~Chatterjee,
Phys. Rev. C \textbf{106} (2022) no.4, 044907
doi:10.1103/PhysRevC.106.044907
[arXiv:2204.02345 [nucl-th]].

\bibitem{STAR:2008jgm}
B.~I.~Abelev \textit{et al.} [STAR],
Phys. Rev. Lett. \textbf{101} (2008), 252301
doi:10.1103/PhysRevLett.101.252301
[arXiv:0807.1518 [nucl-ex]].

\bibitem{ALICE:2013xri}
B.~Abelev \textit{et al.} [ALICE],
Phys. Rev. Lett. \textbf{111} (2013) no.23, 232302
doi:10.1103/PhysRevLett.111.232302
[arXiv:1306.4145 [nucl-ex]].

\bibitem{Parida:2022ppj}
T.~Parida and S.~Chatterjee,
[arXiv:2211.15729 [nucl-th]].

\bibitem{Parida:2025lhn}
T.~Parida,
[arXiv:2505.06522 [nucl-th]].

\bibitem{Parida:2022zse}
T.~Parida and S.~Chatterjee,
[arXiv:2211.15659 [nucl-th]].

\bibitem{NA49:2010lhg}
T.~Anticic \textit{et al.} [NA49],
Phys. Rev. C \textbf{83} (2011), 014901
doi:10.1103/PhysRevC.83.014901
[arXiv:1009.1747 [nucl-ex]].

\bibitem{STAR:2017sal}
L.~Adamczyk \textit{et al.} [STAR],
Phys. Rev. C \textbf{96} (2017) no.4, 044904
doi:10.1103/PhysRevC.96.044904
[arXiv:1701.07065 [nucl-ex]].


\bibitem{STAR:2003cbv}
J.~Adams \textit{et al.} [STAR],
Phys. Rev. C \textbf{71} (2005), 064906
doi:10.1103/PhysRevC.71.064906
[arXiv:nucl-ex/0308033 [nucl-ex]].




\bibitem{CERES:2003sap}
D.~Adamova \textit{et al.} [CERES],
Nucl. Phys. A \textbf{727} (2003), 97-119
doi:10.1016/j.nuclphysa.2003.07.018
[arXiv:nucl-ex/0305002 [nucl-ex]].

\bibitem{PHENIX:2003ccl}
S.~S.~Adler \textit{et al.} [PHENIX],
Phys. Rev. Lett. \textbf{93} (2004), 092301
doi:10.1103/PhysRevLett.93.092301
[arXiv:nucl-ex/0310005 [nucl-ex]].

\bibitem{NA49:2003hxt}
T.~Anticic \textit{et al.} [NA49],
Phys. Rev. C \textbf{70} (2004), 034902
doi:10.1103/PhysRevC.70.034902
[arXiv:hep-ex/0311009 [hep-ex]].


\bibitem{Heckel:2011mtu}
S.~Heckel [ALICE],
J. Phys. G \textbf{38} (2011), 124095
doi:10.1088/0954-3899/38/12/124095
[arXiv:1107.4327 [nucl-ex]].

\bibitem{STAR:2019dow}
J.~Adam \textit{et al.} [STAR],
Phys. Rev. C \textbf{99} (2019) no.4, 044918
doi:10.1103/PhysRevC.99.044918
[arXiv:1901.00837 [nucl-ex]].

\bibitem{STAR:2013ayu}
L.~Adamczyk \textit{et al.} [STAR],
Phys. Rev. C \textbf{88} (2013), 014902
doi:10.1103/PhysRevC.88.014902
[arXiv:1301.2348 [nucl-ex]].


\bibitem{Bozek:2022svy}
P.~Bozek,
Phys. Rev. C \textbf{106} (2022) no.6, L061901
doi:10.1103/PhysRevC.106.L061901
[arXiv:2207.04927 [nucl-th]].

\bibitem{Du:2022yok}
L.~Du, C.~Shen, S.~Jeon and C.~Gale,
Phys. Rev. C \textbf{108} (2023) no.4, L041901
doi:10.1103/PhysRevC.108.L041901
[arXiv:2211.16408 [nucl-th]].

\bibitem{Parida:2023ldu}
T.~Parida and S.~Chatterjee,
[arXiv:2305.08806 [nucl-th]].

\bibitem{Jahan:2025cbp}
S.~A.~Jahan, H.~Roch and C.~Shen,
[arXiv:2507.11394 [nucl-th]].

\bibitem{Parida:2025ddt}
T.~Parida, S.~Chatterjee and S.~Singha,
[arXiv:2503.04660 [nucl-th]].


\bibitem{Broniowski:2016ybk}
W.~Broniowski and P.~Bozek,
Acta Phys. Polon. Supp. \textbf{9} (2016), 189
doi:10.5506/APhysPolBSupp.9.189
[arXiv:1603.08676 [nucl-th]].


\bibitem{Shen:2017bsr}
C.~Shen and B.~Schenke,
Phys. Rev. C \textbf{97} (2018) no.2, 024907
doi:10.1103/PhysRevC.97.024907
[arXiv:1710.00881 [nucl-th]].

\bibitem{Shen:2023awv}
C.~Shen, B.~Schenke and W.~Zhao,
Phys. Rev. Lett. \textbf{132} (2024) no.7, 072301
doi:10.1103/PhysRevLett.132.072301
[arXiv:2310.10787 [nucl-th]].


\bibitem{Du:2018mpf}
L.~Du, U.~Heinz and G.~Vujanovic,
Nucl. Phys. A \textbf{982} (2019), 407-410
doi:10.1016/j.nuclphysa.2018.09.015
[arXiv:1807.04721 [nucl-th]].

\bibitem{Du:2025bhb}
X.~Du, S.~Schlichting and J.~Zhu,
[arXiv:2512.08007 [hep-ph]].

\bibitem{Ryu:2021lnx}
S.~Ryu, V.~Jupic and C.~Shen,
Phys. Rev. C \textbf{104} (2021) no.5, 054908
doi:10.1103/PhysRevC.104.054908
[arXiv:2106.08125 [nucl-th]].


\bibitem{Parida:2026data}
\url{https://doi.org/10.5281/zenodo.19696171}.

\end{thebibliography}
\end{document}